\documentclass[12pt]{article}

\def\lesssim{\mathrel{\mathpalette\vereq<}}
\def\gtrsim{\mathrel{\mathpalette\vereq>}}
\makeatletter
\def\vereq#1#2{\lower3pt\vbox{\baselineskip1.5pt \lineskip1.5pt
\ialign{$\m@th#1\hfill##\hfil$\crcr#2\crcr\sim\crcr}}}
\makeatother

\begin{document}

\begin{titlepage}
\begin{center}
\today     \hfill IASSNS-HEP-99/69 \\
~{} \hfill    LBNL-44042\\
~{} \hfill UCB-PTH-98/29 \\
~{} \hfill hep-ph/9908223\\

\vskip .1in

{\large \bf The Minimal Supersymmetric Leptogenesis}%
\footnote{
This work was supported in part by the
U.S.~Department of Energy under Contracts DE-AC03-76SF00098, in part
by the National Science Foundation under grants PHY-95-13835 and
PHY-95-14797.  TM was also supported by the Marvin L. Goldberger
Membership.  HM was also supported by Alfred P. Sloan Foundation.}

\vskip 0.3in

Takeo Moroi$^{1}$ and Hitoshi Murayama$^{2,3}$

\vskip 0.05in

{\em $^{1}$School of Natural Sciences,
         Institute for Advanced Study\\
         Princeton, New Jersey 08540}

\vskip 0.05in

{\em $^{2}$Theoretical Physics Group\\
     Ernest Orlando Lawrence Berkeley National Laboratory\\
     University of California, Berkeley, California 94720}

\vskip 0.05in

and

\vskip 0.05in

{\em $^{3}$Department of Physics\\
     University of California, Berkeley, California 94720}

\end{center}

\vskip .1in

\begin{abstract}
We reanalyze the supersymmetric leptogenesis model by one of the 
authors (HM) and Yanagida based on $\tilde{L} = H_{u}$ flat direction 
in detail.  We point out that the appropriate amount of baryon 
asymmetry can be generated in this model with the neutrino mass matrix 
consistent with the atmospheric neutrino oscillation and solutions to 
the solar neutrino problem, preferably the small angle MSW solution.  
The reheating temperature can be low enough to avoid the cosmological 
gravitino problem.  This model is the minimal one because it does not 
rely on any new physics beyond supersymmetry and Majorana neutrino 
masses.
\end{abstract}

\end{titlepage}

\newpage

\noindent {\bf 1. Introduction}

Cosmic baryon asymmetry has been one of the biggest puzzles in modern 
cosmology.  It is hoped to understand it in terms of microscopic 
physics and evolution of Universe, if the well-known Sakharov's three 
conditions are satisfied: (1) existence of baryon-number violation, 
(2) existence of CP violation, and (3) departure from thermal 
equilibrium.  A stringent constraint on the baryogenesis models is 
posed by the apparent lack of baryon-number violation.  Recent data 
from Soudan-II and SuperKamiokande experiments \cite{proton} made the 
constraints on proton and neutron decay yet more stronger.  Especially 
given the constraint on $p \rightarrow e^+ \pi^0$, which requires
the mass of the $X$ boson in $SU(5)$ grand unified theories to be 
above $10^{15}$~GeV or so, it is becoming increasingly difficult to 
construct models of baryogenesis with explicit baryon-number 
violation.  On the other hand, the recent evidence for oscillation in 
atmospheric neutrinos from SuperKamiokande \cite{SuperK}, Soudan-II 
\cite{Soudan} and MACRO \cite{MACRO} experiments strongly suggest 
small but finite neutrino masses with $\Delta m^2 = 
10^{-3}$--$10^{-2}$~eV$^2$.  Such small mass scales are unlikely to 
arise for Dirac neutrinos,\footnote{An alternative possibility was 
pointed out recently using large extra dimensions with right-handed 
neutrinos in the bulk \cite{extraD}.} while are naturally understood 
in terms of so-called seesaw mechanism~\cite{seesaw} which in turn 
requires Majorana neutrinos.  Therefore, given the theoretical 
prejudice, there is a strong evidence for lepton-number violation at 
the mass scale of the right-handed neutrinos $M_{R} \sim 
10^{14}$--$10^{15}$~GeV.

Fukugita and Yanagida~\cite{FukYan} suggested that the lepton-number 
violation may be enough for baryogenesis because the electroweak 
anomaly violates $B+L$~\cite{KuzRubSha} and, once lepton number is 
violated, so is baryon number.  If a finite lepton asymmetry is 
created in the Early Universe, it can be partially converted to baryon 
asymmetry without a need for an explicit baryon-number violating 
process which could lead to too-rapid proton decay.  This scenario, 
called leptogenesis, is hence probably the most attractive possibility 
of baryogenesis in the current situations.  The thermally produced 
right-handed neutrinos decay out of equilibrium and the CP-violation 
in the neutrino Yukawa matrix causes a difference between the decay 
rates into leptons and anti-leptons in this model.  The lightest 
right-handed neutrino is typically as heavy as $10^{10}$~GeV in many 
models and therefore the reheating temperature after inflation needs 
to be higher than that (see \cite{recentLG} for a recent discussion).

On the other hand, the large hierarchy between the electroweak scale 
and the scale of lepton number violation $M_{R}$ would require a 
stabilization mechanism for the hierarchy.  Supersymmetry is the best 
solution to this problem currently available.  Therefore it is natural 
and timely to ask what the minimal scheme is for the supersymmetric 
leptogenesis.

Supersymmetry, however, causes a cosmological problem when the 
reheating temperature after the primordial inflation is too high.  
The gravitinos can be produced thermally whose abundance is 
approximately proportional to the reheating temperature.  For a typical 
mass range $m_{3/2} \sim 500$~GeV favored to stabilize the
hierarchy,\footnote{Anomaly mediation of supersymmetry breaking
\cite{AM} may allow much higher gravitino mass keeping most
superparticles light, and hence a higher reheating temperature.} 
the produced gravitinos decay after the Big-Bang Nucleosynthesis and 
they dissociate light elements.  To keep the theory and observation 
consistent, there is an upper bound on the gravitino abundance, and 
hence on the reheating temperature of about $T_{RH} \lesssim 
10^{9}$~GeV for this mass range (for the latest analysis, see 
\cite{gravitino}).  This constraint makes it somewhat difficult to 
produce right-handed neutrino states as required in the original model 
\cite{FukYan}.\footnote{One can get around this problem, however, if
the right-handed neutrinos are produced non-thermally, {\it i.e.}\/,
by the decay of the inflaton \cite{Asaka} or preheating \cite{GPRT}.} 

In this letter, we study the minimal supersymmetric leptogenesis.  Our 
definition of minimal is the particle content of the Minimal 
Supersymmetric Standard Model together with suggested neutrino masses 
as effective dimension-five operators.  This way, we avoid reliance on 
explicit grand-unified models or seesaw models.  The only dependence 
on physics beyond the standard model is supersymmetry and neutrino 
masses.  We show that leptogenesis is possible \`a la 
Affleck--Dine~\cite{AffDin} within this minimal framework as 
originally suggested by one of the authors (HM) and 
Yanagida~\cite{MurYan}, within the current experimental situations.  
The reheating temperature can be low enough to avoid the gravitino 
problem because the model does not require the production of 
right-handed neutrinos.

\noindent {\bf 2. The Model}

We start with the effective dimension-five operators in the 
superpotential
\begin{equation}
        W = \frac{m_{\nu_{i}}}{2 v_{u}^{2}} (L_{i} H_{u}) (L_{i} H_{u})
        = \frac{1}{2M_{i}} (L_{i} H_{u}) (L_{i} H_{u})
\label{eq:W}
\end{equation}
which are necessary to incorporate finite neutrino masses (and hence 
lepton-number violation) to the Minimal Supersymmetric Standard Model.  
The vacuum expectation value for $H_u$ is $v_u = \langle H_u \rangle
= (174$~GeV) $\sin\beta$ and we assume a moderately large $\tan\beta
\gtrsim 5$ throughout the paper.\footnote{For smaller $\tan\beta
\gtrsim 1.4$ allowed by the perturbative top Yukawa coupling,
the resulting lepton asymmetry changes only up to a factor of two.}
The index $i$ runs over three mass eigenstates of neutrinos.  It will 
turn out that the relevant direction is most likely the lightest 
neutrino as we will see later and henceforth suppress the index $i$.  
Along the flat direction $\tilde{L} = H_{u} = \phi/\sqrt{2}$, (the factor 
$\sqrt{2}$ is necessary to ensure the canonical kinetic term: $L^{*} L 
+ H_{u}^{*} H_{u} = \phi^{*} \phi$) the potential is given by
\begin{equation}
        V = m^{2} |\phi|^{2} + \frac{A}{8M} (\phi^{4}+\phi^{*4})
        + \frac{1}{4M^{2}}|\phi|^{6} .
        \label{eq:pontential}
\end{equation}
 Here, the first two terms are from SUSY breaking and we assume $m\sim
A\sim {\rm 100~GeV-1~TeV}$, while the last term is from superpotential
given in Eq.~(\ref{eq:W}).  The mass scale $M$ does not necessarily be
the mass scale of right-handed neutrinos; it can actually be much
higher.  Our philosophy is not to discuss the precise origin of such
operator, but rather use this effective operator only.

\noindent {\bf 3. Initial Amplitude}

Since this direction is $D$-flat and also $F$-flat in the $m_\nu 
\rightarrow 0$ limit, it can have a large amplitude at the end of the 
primordial inflation.  It turns out that the resulting baryon 
asymmetry does not depend on the precise value of the initial 
amplitude.  All we need is that the initial amplitude is larger 
than $\sqrt{m M}$ as we will see later.  We would like to discuss various 
mechanisms to generate a large amplitude proposed in the literature.  
We have the expansion rate of the Universe during the primordial 
inflation of $H_{\it inf} \sim 10^{12}$--$10^{14}$~GeV as suggested by 
popular models~\cite{inflation}.

To discuss the amplitude in the Early Universe, one needs to pay
special attention to the fact that the expanding Universe itself
breaks global supersymmetry and may modify the scalar potential
\cite{DinRanTho}.  The explicit form of the supersymmetry breaking,
however, depends on the details of the K\"ahler potential and hence is
model dependent.  Therefore we consider three possibilities in this
letter.  (1) No modification from the flat-space potential, which
occurs at the tree-level of the no-scale type supergravity
\cite{GaiMurOli}, (2) loop-level modifications from the flat-space
potential \cite{GaiMurOli}, and (3) tree-level modifications from the
flat-space potential \cite{DinRanTho}.  In all cases, we find it is
easy to have an initial amplitude necessary for the model: $\phi_{\it
inf} \gg \sqrt{m M}$ as we will see later.

In case (1), one idea to generate a large initial amplitude is to 
start with a ``chaotic'' initial condition \cite{chaotic},\footnote{We 
do not need $V \sim M_{\rm Planck}^{4}$ as suggested in
\cite{chaotic}, however.} where all 
scalar fields start with large values.  During the course of 
inflation, most of the fields are diluted exponentially due to their 
potential, but some remain if their potential is flat enough.  Along 
our flat direction, the potential is eventually dominated by 
$|\phi|^6$ term once the amplitude becomes small enough whatever the 
physics at high energy scale $\gg M$ is.  The evolution is given by 
the slow-rolling approximation, and we can solve the equation of 
motion
\begin{equation}
  3 H_{\it inf} \dot{\phi} + \frac{3}{4M^{2}} |\phi|^{4} \phi = 0.
\end{equation}
For a more-or-less constant expansion rate $H=H_{\it inf}$ during the
inflation, its solution is
\begin{equation}
  \frac{1}{|\phi|^{4}(t)} - \frac{1}{|\phi|^{4}(0)}
  = \frac{1}{M^{2} H_{\it inf}} t .
\end{equation}
Therefore with an $e$-folding of $N=H_{\it inf}t$,
\begin{equation}
  |\phi|^{4}(t) = \frac{1}{|\phi|^{-4}(0) + N/(M^{2} H_{\it 
inf}^{2})}.
\end{equation}
Clearly, the amplitude at the end of inflation is almost predicted as
long as $|\phi(0)|$ is large enough,
\begin{equation}
  |\phi|^{4}(t) = \frac{M^{2} H_{\it inf}^{2}}{N}.
\end{equation}
The $e$-folding $N$ required to solve the horizon and flatness 
problems is $N \gtrsim 100$ \cite{Lin}.

The cases (2) and (3) use a possible negative mass squared of 
order $-H_{\it inf}^{2}/16\pi^{2}$ \cite{GaiMurOli} or $-H_{\it 
inf}^{2}$ \cite{DinRanTho}.  Minimizing
\begin{equation}
  V = - c H_{\it inf}^{2} |\phi|^{2} + \frac{1}{4M^{2}}|\phi|^{6},
\end{equation}
with $c\sim O(1)$ or $O(1/16\pi^{2})$, we find $-c H_{\it inf}^{2} 
\phi + (3/4M^{2}) |\phi|^{4} \phi =0$, or $|\phi|^{4} = 4 c H_{\it 
inf}^{2} M^{2}/3$.

Both above arguments are classical.  One may worry that the quantum 
fluctuation during the de Sitter expansion of Universe may modify the 
estimate.  An effective Hawking--Gibbons temperature is given by 
$T_{HG} = H_{\it inf}/2\pi$ and the fields acquire expectation values 
which correspond to the potential energy density of the order of the 
$T_{HG}^{4}$ \cite{Lin}.  This effect suggests $|\phi|^{6} \sim M^{2} 
H_{\it inf}^{4}$, which is typically smaller than the other estimates 
for the range of $M$ of our interest (see below).  This smallness of 
the quantum fluctuation therefore justifies the classical treatment 
above for the flat direction because the fluctuation around the 
classical value can be neglected.

Based on the above discussions, we take an order of magnitude
estimate of the amplitude of $\phi$ at the end of the inflation 
$t_{\it inf}$:
\begin{eqnarray}
        \phi_{inf}^4 \equiv |\phi(t_{\it inf})|^{4} \sim M^{2} H_{\it inf}^{2},
\end{eqnarray}
or slightly smaller for the rest of the analysis.  The result turns 
out not to depend on the precise value of the initial amplitude; the 
only requirement is that $\phi_{\it inf} \gg \sqrt{m M}$.

\noindent {\bf 4. Evolution}

Given the above initial condition, we follow the cosmological 
evolution after the end of the inflation.  While the inflaton 
oscillates around the minimum of its potential, the flat direction 
$\phi$ gets diluted.  For case (1), the equation of motion of the 
$\phi$ when the potential is dominated by $\phi^{6}$ term is given by
\begin{equation}
        \ddot{\phi} + \frac{2}{t} \dot{\phi} + \frac{3}{4} 
        \frac{\phi^{5}}{M^{2}} = 0,
        \label{eq:eom}
\end{equation}
assuming the matter dominated expansion $H = 2/3t$.  For case (2), 
there is an additional term of order $-\frac{1}{16\pi^{2} t^{2}} \phi$ 
in the l.h.s. of the equation, and similarly $-\frac{1}{t^{2}} \phi$ 
for case (3).  Despite the apparent difference between three cases, 
$\phi$ turns out to track $\sqrt{M H}$ until the expansion rate slows 
down to $H \sim m$ as we will explain below.

We first discuss case (1) with flat-space potential because of its 
simplicity.  The behavior of $|\phi|$ depends on which term dominates 
its potential, since the potential changes its shape at
\begin{equation}
  |\phi| \sim \sqrt{mM} \equiv \phi_{LG}.
\end{equation}
At the time when $|\phi| \sim \phi_{LG}$, the lepton number gets fixed 
(we refer to this point as LG, for LeptoGenesis).  For $|\phi|\gtrsim 
\phi_{LG}$, $|\phi|^6$ term dominates the potential and the equation 
of motion (\ref{eq:eom}) should be used.  Using the slow-roll 
approximation $|\ddot{\phi}| \ll |2\dot{\phi}/t|$, the solution is 
given by
\begin{equation}
        \phi (t)^{4} = \left[ \phi_{\it inf}^{-4} 
        + 3\frac{t^{2}-t_{\it inf}^{2}}{4M^2} \right]^{-1},
        \label{eq:slowroll}
\end{equation}
and it is easy to check that the motion is critically damped when 
$\phi_{\it inf} \lesssim \sqrt{M H}$.  Therefore, independent of the 
precise initial value of $\phi_{\it inf}$, the motion starts only when 
$\phi_{\it inf} \sim \sqrt{M H}$.  After $\phi$ starts to move, one 
can use two extreme approximations to study the dilution of $\phi$: 
the slow-roll approximation and the virial motion.  With the slow-roll 
approximation, the solution (\ref{eq:slowroll}) above dilutes as $\phi^{4} 
\propto R^{-3} \propto H^{2}$.  The opposite limit of the virial 
(oscillatory) motion with $\phi^{6}$ potential gives $\dot{\phi}^{2}/2 
= 3 V$.  Then the equation of motion $\ddot{\phi} + 3 H \dot{\phi} + 
V'=0$ can be rewritten in terms of the energy density $\rho_{\phi} = 
\dot{\phi}^{2}/2 + V$ as $\dot{\rho_{\phi}} = - 3 H \dot{\phi}^{2} = - 
\frac{9}{2} H \rho_{\phi}$, and hence $\rho_{\phi} \propto R^{-9/2}$.  
Therefore we again obtain $\phi^{4} \propto R^{-3} \propto H^{2}$.  
The true motion is somewhere between the slow-roll and virial limits 
but the agreement of both extremes suggests that this is the true 
power law of dilution.  Then the amplitude $\phi$ always traces 
$\sqrt{M H}$.  On the other hand, once $|\phi|\lesssim \phi_{LG}$, 
$\phi$ has a quadratic potential.  In this case, the particle picture 
is valid for $\phi$ and $\phi$ dilutes as $|\phi|^2\propto R^{-3}$.

For a low reheating temperature ($T_{RH} \lesssim 10^{10}$~GeV), 
$|\phi|\sim \phi_{LG}$ is realized in the inflaton-dominated Universe.  
Because $\phi$ tracks $\sqrt{MH}$, this occurs when $H \sim H_{LG} = 
m$.  This corresponds to the inflaton energy density
\begin{equation}
        \rho_\psi (t_{LG}) \sim 3 m^2 M_*^2 ,
\end{equation}
where $M_{*} = M_{\rm Planck}/\sqrt{8\pi} = 2.4 \times 10^{18}$~GeV is 
the reduced Planck mass.

Lepton number density at this time is easily estimated.  Define the 
$\phi$-number density $n_\phi\equiv i(\dot{\phi}^*\phi - 
\phi^*\dot{\phi})$, which is related to the lepton number density as 
$n_L = \frac{1}{2}n_\phi$, then its equation of motion is given by
\begin{equation}
        \dot{n}_\phi + 3 H n_\phi
        = i\left( \frac{\partial V(\phi)}{\partial \phi} - {\rm h.c.} \right)
        = \frac{A}{M} {\rm Im}(\phi^{4}).
\end{equation}
As a result, we find that
\begin{equation}
        \frac{\partial}{\partial t}
        (R^{3} n_\phi) = R^{3} \frac{A}{M} {\rm Im} (\phi^{4}),
\end{equation}
and hence
\begin{equation}
        (R^{3} n_\phi) (t_{LG}) = \int_{0}^{\infty} dt\, R^{3} \frac{A}{M} 
        {\rm Im} (\phi^{4}).
        \label{eq:dot(R^3n)}
\end{equation}
For $t < t_{LG}$ in the integral, $\phi^{6}$ dominates the potential 
and $\phi^4\propto R^{-3}$.  The integral for this time range
just gives the time interval 
$t_{LG}$.  For $t > t_{LG}$, $\phi^{2}$ dominates the potential 
and $\phi^{4}\propto R^{-6}$.  The integral for this time range gives the 
same contribution as the previous one.  
Therefore the resulting lepton asymmetry is dominated by 
the late time contribution at $H \sim m$ and rather insensitive to the 
history before then.  For matter-dominated Universe, 
$t_{LG}=\frac{2}{3}H_{LG}^{-1}$, and hence
\begin{equation}
        n_\phi (t_{LG}) \sim \frac{4A}{3mM} {\rm Im} (\phi^4).
\end{equation}
By substituting $\phi^4 \sim m^2M^2e^{4i\theta}$ and $A \sim m$,
\begin{equation}
        n_\phi (t_{LG}) \sim \frac{4}{3} m^{2} M \sin 4\theta.
\end{equation}
The size of the resulting lepton number is determined by the initial 
phase of the field $\theta$.  This is because the potential has 
``valleys'' along $\theta = \pi/4, 3\pi/4, 5\pi/4, 7\pi/4$ and ``ridges'' 
along $\theta = 0, \pi/2, \pi, 3\pi/2$ and the initial phase 
generates a rotational motion of $\phi$ on the complex plane when it 
rolls down the ``slope'' from the ``ridge.''  In this sense, the 
lepton number is determined by a ``spontaneous'' CP violation because 
of the random phase the field acquired during the inflation.

When the inflaton decays, the Universe is reheated.  With the 
instantaneous decay approximation, $\rho_\psi 
(t_{RH})=\frac{\pi^2}{30}g_*T_{RH}^4$, and we obtain
\begin{equation}
        \left( \frac{R_{RH}}{R_{LG}} \right)^3
        = \frac{\rho_\psi (t_{LG})}{\rho_\psi (t_{RH})}
        \sim \frac{3m^2M_*^2}{\frac{\pi^2}{30}g_*T_{RH}^4}.
\end{equation}
Therefore,  the $\phi$ number density at the reheating is given by
\begin{equation}
        n_\phi(t_{RH})
        = n_\phi(t_{LG}) \left( \frac{R_{LG}}{R_{RH}} \right)^3
        \sim \frac{4M}{9M_*^2} \sin 4\theta \times \frac{\pi^2}{30}g_* T_{RH}^4,
\end{equation}
or using $s(t_{RH})=\frac{2\pi^2}{45}g_*T_{RH}^3$, we obtain
\begin{equation}
        \frac{n_\phi}{s} \sim \frac{MT_{RH}}{3M_*^2} \sin 4\theta.
        \label{eq:n_phi/s}
\end{equation}

The above estimate is valid as long as $|\phi|\sim \phi_{LG}$ is 
realized in the inflaton-dominated Universe.  This is the case if 
$3m^2M_*^2\gtrsim \frac{\pi^2}{30}g_*T_{RH}^4$, or equivalently, 
$T_{RH}\lesssim 10^{10}~{\rm GeV}$ for $m\sim {\rm 100~GeV}$.  For 
higher reheating temperature, however, $|\phi|\sim \phi_{LG}$ happens 
after the reheating.  At the end of the inflation, $H = H_{\it inf}$, 
and the energy density of the oscillating inflaton $\psi$ is 
$\rho_{\psi} = 3 H_{\it inf}^{2} M_{*}^{2}$.  Using 
$\rho_{\psi}(t_{RH}) =\frac{\pi^{2}}{30}g_* T_{RH}^{4}$, we find 
$(R_{\it inf}/R_{RH})^{3}\sim (\frac{\pi^{2}}{30} T_{RH}^{4}) / (3 
H_{\it inf}^{2} M_{*}^{2})$.  At the same time, the flat direction is 
diluted as
\begin{equation}
        |\phi|^{4}(t_{RH})
        \sim M^{2} H_{\it inf}^{2} \left(\frac{R_{\it inf}}{R_{RH}}\right)^3
        \sim \frac{M^{2}}{3M_{*}^{2}} \times \frac{\pi^{2}}{30} g_*T_{RH}^4,
\end{equation}
which is still larger than $\phi_{LG}^4$ for $T_{RH}\gtrsim
10^{10}~{\rm GeV}$.

At the time when $|\phi|^{4}\sim m^{2} M^{2}$, the lepton number gets
fixed.  The scale factor at this point is determined by
\begin{equation}
        \left(\frac{R_{RH}}{R_{LG}}\right)^{3}
        = \left(\frac{\phi_{LG}}{\phi_{RH}}\right)^4
        \sim \frac{3 m^{2}M_{*}^{2}}{\frac{\pi^{2}}{30}g_* T_{RH}^{4}}.
\end{equation}
The radiation energy density at $t=t_{LG}$ is therefore given by
\begin{equation}
        \rho_{rad} (t_{LG})
        = \left(\frac{\pi^{2}}{30}g_* T_{RH}^{4}\right)
        \left(\frac{R_{RH}}{R_{LG}}\right)^{4}
        \sim 3 m^{2} M_{*}^{2} \left(\frac{3 m^{2} M_{*}^{2}}
        {\frac{\pi^{2}}{30} g_* T_{RH}^{4}}\right)^{1/3}.
\end{equation}
This gives the expansion rate at this point
\begin{equation}
        H_{LG}\sim m \left(\frac{3
        m^{2} M_{*}^{2}}{\frac{\pi^{2}}{30} g_*T_{RH}^{4}}\right)^{1/6}.
\end{equation}
The evolution of $n_\phi$ is solved as in the previous case.  For 
radiation-dominated Universe, $t_{LG} = \frac{1}{2}H_{LG}$, and hence 
Eq.~(\ref{eq:dot(R^3n)}) gives
\begin{equation}
        n_\phi (t_{LG}) \sim \frac{1}{2} m^{2}
        M\left(\frac{3 m^{2} M_{*}^{2}}
        {\frac{\pi^{2}}{30} g_*T_{RH}^{4}}\right)^{-1/6} \sin 4\theta,
\end{equation}
where we used $\phi^{4} \sim m^{2} M^{2} e^{4i\theta}$ and $A\sim m$.  
If we naively adopt the above relation, $n_\phi$ is bigger than 
$m^2M\sim m|\phi_{LG}|^2$ for $\theta\sim O(1)$.  However, at this 
point, the $\phi$ motion allows a particle interpretation because the 
potential is dominated by the quadratic piece, and the maximum 
possible number of particles is given by $\sim m |\phi|^{2}$.  If we 
take the above estimate literally, $n_\phi$ may be bigger than $m 
|\phi|^{2}$: a contradiction.  Then the conclusion is that $n_\phi$ 
has already saturated before this stage.  Therefore, some of the 
previous discussions for $T_{RH}\gtrsim (mM_*)^{1/2}$ are somewhat 
irrelevant, and we can simply estimate $n_\phi$ at $\phi\sim\phi_{LG}$ 
to be $\sim m^{2} M$.  Then, using the entropy at $t_{LG}$:
\begin{equation}
        s(t_{LG})
        = s(t_{RH}) \left(\frac{R_{RH}}{R_{LG}}\right)^{3}
        \sim \frac{4 m^{2} M_{*}^{2}}{T_{RH}},
\end{equation}
we obtain a similar $\phi$-to-entropy ratio as given in
Eq.~(\ref{eq:n_phi/s}) up to an $O(1)$ coefficient.  We use the result
given in Eq.~(\ref{eq:n_phi/s}) for the numerical estimation below.

With the estimated $n_\phi/s$ (\ref{eq:n_phi/s}), 
the lepton-to-entropy ratio is given by
\begin{equation}
        Y_{L} \equiv \frac{n_{L}}{s} = \frac{n_\phi}{2s}
        \sim \frac{1}{6} \frac{M T_{RH}}{M_{*}^{2}} \sin 4\theta
        = \frac{1}{6} \frac{v_{u}^{2} T_{RH}}{m_{\nu} M_{*}^{2}} \sin 4\theta.
        \label{eq:YL}
\end{equation}
Since the lepton number is inversely 
proportional to the mass of the neutrino, it is dominated by the 
lightest mass eigenvalue $m_{\nu_{1}}$, possibly mostly the electron 
neutrino state.

So far the discussion used only the flat-space potential (case (1)).  
For case (2), the discussion is modified only slightly.  The field 
amplitude dilutes as $\phi \sim \sqrt{M H}$ because the negative mass 
squared $-H^{2}/16\pi^{2}$ is smaller than the expansion rate and 
hence does not modify the discussion with flat-space potential.  The 
main issue here is if the phase direction of the field gets settled to 
the ``valley.''  Because of loop-suppressed effect of the expansion in 
the potential, the potential along the phase direction is given by $((A 
+ c H/16\pi^{2})\phi^{4} + c.c.)/8M$ with $c = O(1)$, and the 
curvature along the phase direction is of the order of $c H 
\phi^{2}/16\pi^{2} M \sim c H^{2}/16\pi^{2}$.  This is always smaller 
than the expansion rate and hence the motion along the phase direction 
is critically damped.  Therefore the initial phase of the potential is 
kept intact until the expansion slows down to $H \sim m$ when the 
flat-space potential is recovered and the field starts rolling down 
the slope from the ``ridge.''  The estimate of the resulting lepton 
asymmetry is hence the same as in the case (1) with the flat-space 
potential.

The discussion is different for case (3) where the effect of the 
expansion introduces a large supersymmetry breaking effect of $O(H)$ 
in the potential.  This case was discussed in detail in \cite{DRT2}, 
and we summarize the situation briefly here.  The field amplitude 
always tracks the minimum of the potential $\phi \sim \sqrt{M H}$ 
because the curvature $-H^{2} |\phi|^{2}$ is of the same order of 
magnitude as the expansion rate itself and field always rolls down to 
the minimum at any given time.  Therefore the field amplitude is the 
same as other two cases even though the reason for it is quite 
different.  Similarly, the potential along the phase direction $((A + c 
H)\phi^{4} + c.c.)/8M$ with $c = O(1)$ produces the curvature 
along the phase direction is also of the order of $c H \phi^{2}/ M 
\sim H^{2}$ and hence the field settles to the ``valley'' and tracks 
it.  However, the potential can generate a rotational motion when $H$ 
becomes smaller than $A$ because of a possible relative phase between 
two terms $A$ and $cH$.  The phase for the ``valleys'' quickly shifts to that of 
the flat-space potential at this moment and the field which 
tracked the valley with the $H$ effect starts moving to the new valley 
with the $A$ term only.  The estimate of the rotational motion is 
basically the same as the earlier two cases and hence the resulting 
lepton asymmetry as well.  What is different from the previous cases 
is that the CP violation is in the relative phase between $A$ and 
$cH$ and originates from the microscopic physics rather than a 
spontaneous random phase.

To summarize, in all cases (1), (2), (3) considered, the final lepton 
asymmetry from the motion of the $D$-flat direction $\tilde{L}=H_{u}$ is given 
by Eq.~(\ref{eq:YL}) which confirms the original estimate in 
\cite{MurYan}.  Therefore the result is quite robust and model 
independent.

\noindent {\bf 5.  Estimate of Baryon Asymmetry}

In order to proceed to a numerical estimate of the generated lepton 
asymmetry, we need to have some idea on the lightest neutrino mass 
from the currently available data.  Here we have to rely on certain 
assumptions.  

We assume hierarchical neutrino masses similar to quark and charged 
lepton masses, and hence that $\Delta m^{2}$ from neutrino oscillation 
give larger of the mass eigenvalues between two neutrino states 
relevant to the particular oscillation mode.  The atmospheric neutrino 
data require that the largest two eigenvalues have $\Delta m^{2} = 
10^{-3}$--$10^{-2}$~eV$^{2}$, and hence according to our assumption, 
$m_{\nu_{3}} = 0.03$--0.1~eV. The small angle MSW solution suggests 
$\Delta m^{2} \simeq 4$--$10 \times 10^{-6}$~eV$^{2}$, which should give the 
second lightest mass eigenvalue $m_{\nu_{2}} \simeq 0.0020$--0.0032~eV and the 
mixing angle $\sin^{2} 2 \theta_{MNS} \simeq 1 \times 10^{-3}$--$1 
\times 10^{-2}$
\cite{solar}.  The estimate of the lightest mass eigenvalue depends on 
the assumptions on the mass matrix.  If the MNS 
(Maki--Nakagawa--Sakata) mixing angle \cite{MNS} is solely due to the 
neutrino Majorana matrix, the lightest eigenvalue is likely to be 
around $m_{\nu_{1}} \sim m_{\nu_{2}} \theta_{MNS}^{2} \sim (0.5$--$8) \times 
10^{-6}$~eV.
This approximate relation between the mass eigenvalue and the angle is 
quite generic unless most of the mixing is attributed to the charged 
lepton sector.  For instance the models based on coset-space family 
unification on $E_{7}/SU(5) \times U(1)^{3}$ \cite{YS}, 
string-inspired anomalous $U(1)$ \cite{ILR}, and $SU(5)$ grand unified 
model with Abelian flavor symmetry \cite{AF} all give this approximate 
relation.  However if the mixing angle is coming from the charged 
lepton sector, the mass eigenvalue can be smaller.  

For the large angle MSW solution, $m_{\nu_{2}} \simeq 0.003$--0.01~eV, 
and the lightest eigenvalue is likely to be not much smaller than 
$m_{\nu_{2}}$ if the large MNS angle comes from the neutrino mass 
matrix, but can be much smaller if the MNS angle originates in the 
charged lepton mass matrix.  This might be considered a fine-tuning 
however because the determinant of the mass matrix (either charged 
lepton or neutrino) should be much smaller than the typical size of 
the elements.  The vacuum solution most likely gives $m_{\nu_{1}} 
\simeq m_{\nu_{2}} \simeq 10^{-5}$~eV, which may also need a 
fine-tuning in the mass matrices.  For the purpose of the estimation 
below, we take $m_{\nu_{1}} \sim 2.5 \times 10^{-6}$~eV, preferred by 
the small angle MSW solution, but the result can be easily scaled 
according to the value of $m_{\nu_{1}}$.

The reheating temperature cannot be arbitrarily high because of the
cosmological constraints on the gravitino production.  For the
range of gravitino mass usually considered, $m_{3/2} \simeq 500$~GeV
or so, the primordial gravitinos will decay with the lifetime of
$\tau_{3/2} \simeq 2 \times 10^{5}$~sec and therefore can potentially
destroy the light elements synthesized by Big-Bang Nucleosynthesis.
The most robust constraint is from the thermally produced gravitino,
which requires the reheating temperature to be less than $T_{RH}
\lesssim 10^{9}$~GeV \cite{gravitino}.  Another constraint is obtained
from the non-thermal production after the inflationary epoch, which
may give a more stringent constraint on the reheating temperature
\cite{GTR}.  However, the number density of the non-thermally produced
gravitino is model-dependent.  For example, for the chaotic inflation
model with the inflaton superpotential of $W_{\it inflaton}\sim
\frac{1}{2}M_{\it inflaton}\phi_{\it inflaton}^2$, the gravitino to entropy
ratio is of order $M_{\it inflaton}T_{RH}/M_*^2$, which is smaller than
that of the thermally produced gravitino.  Because of the
model-dependence, we do not include the constraint from the
non-thermally produced gravitino into our discussion, and we take the
canonical value of $T_{RH}=3 \times 10^{8}$~GeV for the estimation
below.%
\footnote{Moduli fields may also be dangerous.  In particular, if they
have large initial amplitudes of $O(M_*)$, their primordial number
density becomes so large that the Big-Bang Nucleosynthesis is
seriously damaged.  The initial amplitude may be suppressed, for
example, if the minimum of the moduli potential is given by a symmetry
enhanced point~\cite{DinRanTho}.  Non-thermal production of the moduli
may be also important, but is also model-dependent~\cite{GTR}.  In
particular, if the modulus have a large effective mass of $O(H_{\it inf})$
during the inflation, or if the modulus is conformally coupled, then
its non-thermal production after the inflation is not effective.}

Now come the numerical estimates.  We find the lepton asymmetry
(\ref{eq:YL}) to be
\begin{equation}
        Y_{L} \sim 1.1 \times 10^{-10}
        \left(\frac{2.5 \times 10^{-6}~\mbox{eV}}{m_{\nu_{1}}}\right)
        \left(\frac{T_{RH}}{3\times 10^{8}~\mbox{GeV}}\right) \sin 4\theta.
\end{equation}
From the chemical equilibrium between lepton and baryon numbers, we 
obtain $Y_{B} = 0.35 Y_{L}$ \cite{KS}, while 
the number density of photons now is related to the entropy density by 
$n_{\gamma}/s = (410.89~\mbox{cm}^{-3})/(2892.4~\mbox{cm}^{-3})$, and 
in the end we find
\begin{equation}
        \eta = \frac{n_{B}}{n_{\gamma}} \simeq 2.46 Y_{L}
        \sim 2.6 \times
        10^{-10} \left(\frac{2.5 \times 10^{-6}~\mbox{eV}}{m_{\nu_{1}}}\right)
        \left(\frac{T_{RH}}{3 \times 10^{8}~\mbox{GeV}}\right) \sin 4\theta.
\end{equation}
Recall that we used the preferred range by the small angle MSW 
solution $m_{\nu_{1}} \sim (0.5$--$8) \times 10^{-6}$~eV.
This estimate should be compared to the value determined from the 
Big-Bang Nucleosynthesis $\eta\simeq (4-6)\times 10^{-10}$.  Therefore 
this mechanism can generate the cosmic baryon asymmetry of the correct 
order of magnitude without conflicting the constraints on the 
cosmological gravitino production within the range of neutrino mass 
preferred by the atmospheric 
neutrino data and the small angle MSW solution to the solar neutrino 
problem.  However, one should note that such a small neutrino mass can 
be accommodated even with other solutions to the solar neutrino problem 
({\it i.e.}\/, large angle MSW and vacuum oscillation solutions) by 
attributing the MNS angle to the charged lepton sector or by a 
fine tuning in the neutrino mass matrix.

\noindent {\bf 6. Conclusion}

To conclude, we have shown that the minimal supersymmetric standard 
model together with Majorana mass of neutrinos can generate the cosmic 
baryon asymmetry without any additional new physics, hence the minimal 
supersymmetric leptogenesis.  The required neutrino mass is consistent 
with the mass matrices proposed to explain the atmospheric as well as 
solar neutrino data with the small angle MSW solution.  Other 
solutions to the solar neutrino problem can also be accommodated if the 
lightest neutrino mass eigenvalue is of the order of $10^{-6}$~eV. The 
reheating temperature can be low enough to avoid cosmological 
gravitino problem.

\noindent{\bf Acknowledgements}

We thank Steve Martin and Pierre Ramond for helpful conversations, and
Gian Giudice and Antonio Riotto for useful comments.  HM also thanks
Masahiro Kawasaki and Tsutomu Yanagida for encouraging him to write up
this work.  We would like to thank Aspen Center for Physics where this
work was completed.  This work was supported in part by the
U.S.~Department of Energy under Contracts DE-AC03-76SF00098, and in
part by the National Science Foundation under grants PHY-95-13835 and
PHY-95-14797.  TM was also supported by the Marvin L. Goldberger
Membership.  HM was also supported by Alfred P. Sloan Foundation.

\end{document}